\documentclass[twocolumn,pre,showpacs]{revtex4}
\usepackage{epsf,graphicx,amssymb}

\begin{document}
\title{Wave-vortex interaction}
\author{Claudio Falc\'on$^{1,2}$}
\author{St\'ephan Fauve$^{1}$}
\address{$^{1}$Laboratoire de Physique Statistique, \'Ecole Normale Sup\'erieure, CNRS , 24, rue Lhomond, 75 005 Paris, France \\ $^{2}$Departamento de F\'isica, Facultad de Ciencias F\'isicas y Matem\'aticas, Universidad de Chile, Casilla 487-3, Santiago, Chile}
\date{\today}

\begin{abstract}

We present an experimental study on the effect of a electromagneticaly generated vortex flow on parametrically amplified waves at the surface of a fluid. The underlying vortex flow, generated by a periodic Lorentz force, creates spatio-temporal fluctuations that interact nonlinearly with the standing surface waves. We characterize the bifurcation diagram and measure the power spectrum density (PSD) of the local surface wave amplitude. We show that the parametric instability threshold increases with increasing intensity of the vortex flow. 

\end{abstract}
\date{\today}
\maketitle

\section{Introduction}

Wave motion on fluid interfaces is affected by the fluid's bulk motion and its underlying flow structure\cite{Whitham}. Surface waves of wavelength $\lambda$ and phase velocity $c_{\lambda}$ can experience advection and eigenfrequency shifts due to the presence of a mean flow, thus changing the behavior and properties of wave patterns. In the case of strongly fluctuating or even turbulent flows the degree of complexity of wave propagation increases. Wave motion on the surface of a turbulent fluid has been a problem of interest since the early works of Phillips\cite{Phillips58} on the scattering of a gravity surface wave of wavelength $\lambda$ and phase velocity $c_{\lambda}=\sqrt{ g\lambda/2\pi}$, by turbulent velocity fluctuations $u_t$ in the limit $u_t\ll c_{\lambda}$. He pointed out the possibility of wave generation and wave dissipation induced by turbulent fluctuations, which has been experimentally studied on surface waves\cite{Nature}. The aim of our work is to characterize the later effect on an experimental system, where a wave pattern of a given amplitude and wavelength develops over an externally controlled fluctuating background. To that end we have studied the effect of an electromagnetically generated vortex flow on parametrically amplified surface waves. 

It was shown by Faraday\cite{Faraday} that surface waves can be excited on a layer of fluid by means of parametric resonance\cite{Parametric}. By vibrating periodically the fluid container at a given frequency, he observed a pattern oscillating at half the forcing frequency over the surface of the fluid, above a threshold value of the control parameter (the vibration amplitude). The effect of fluctuations on parametric instabilities (the modulation of certain parameters of the system) has been studied theoretically\cite{Stratonovich} and experimentally\cite{Fauve,Chinitos,Otros}. In all of these cases, the source of fluctuations has been either spatial or temporal. To our knowledge, there is no experimental study where controlled spatio-temporal fluctuations are present in the cases of unstable modes developing through parametric instabilites. Here, parametrically amplified gravity surface waves are randomly forced by means of a periodic vortex flow. This fluctuating background flow is generated by a periodic Lorentz force acting on the fluid which supports the gravity surface waves. In our experimental set-up, the electromagnetically induced velocity fluctuations of the vortex flow $\sigma(v)$ are small with respect to $c_{\lambda}$, as in \cite{Phillips58}: the Froude number Fr = $\sigma(v)^{2}/c_{\lambda}^{2}\simeq$ is small (Fr$\simeq$ 0.05). The typical surface energy of the parametrically amplified gravity waves is largely superior to the kinetic energy of fluctuating flow, as it is shown by the system's Weber number We = $\rho \sigma(v)^{2} \lambda/\gamma\simeq$ 0.05, where $\rho$ and $\gamma$ are the fluid's density and surface tension. Although we are in low We and Fr limit, the effect of the vortex flow fluctuations on the properties of parametrically amplified waves is not negligible. 

In this paper, we show that the action of a fluctuating velocity field on parametrically amplified gravity surface waves can inhibit the growth of the standing wave pattern, thus increasing the instability threshold above its deterministic value. We charaterize this effect by locally measuring the velocity field $v_1$ generated by a periodic Lorentz force and the local wave amplitude $h_1$ of the surface waves. The manuscript is organized as follows. In Section II we present the experimental set-up and measurement techniques used to study the local fields of interest. In Section III, we study separatedly the effect of the parametric excitation and the Lorentz force on the fluctuations of surface waves. The combined effect of both excitation mechanisms is studied in Section IV. We show that the parametric instability threshold increases when the underlying vortex flow is active. Finally, in Section V we present the conclusions and perspectives of this work.

\section{Experimental setup and measurement techinques}
\begin{figure}[h]
 \centering
 \includegraphics[width=\columnwidth]{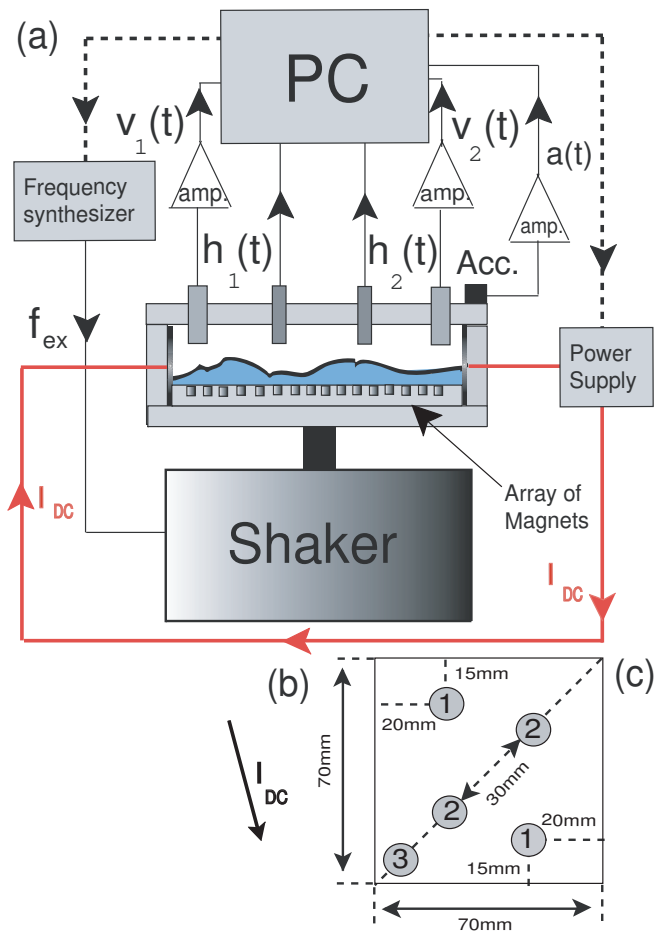}
 \caption{(color online) a) Experimental set-up. b) Bottom of the experimental cell showing the hexagonal array of alternating polarity magnets, used to generate the periodic Lorentz force ${\bf F}_L$. Black arrow shows the sense of the applied DC current. c) Top view of the cell showing sensor placement: 1) Viv\`es probes, 2) inductive sensors, 3) accelerometer. }
 \label{Fig1}
\end{figure}

The experimental set-up is displayed in Fig.~\ref{Fig1}. A Plexiglass container of 70 x 70 mm$^{2}$ is filled with mercury (density $\rho$ = 13.6 x 10$^{4}$ kg/m$^{3}$, kinematic viscosity $\nu$= 1.2 x 10$^{-7}$ m$^{2}$/s and surface tension $\gamma$= 0.4 N/m) up to 5 mm. At the bottom of the cell, alternating vertical polarity magnets (diameter $\phi$=5 mm, height $h=$ 8 mm) made of neodymium and nickel-coated are placed with a 1 mm gap between them in an hexagonal array (wavelength $\lambda$=6 mm). The magnetic field strength at the surface of the fluid on top of a magnet is $500$ G. Two nickel-barnished copper electrodes are glued at opposite sides of the cell. A fine layer of Ni is deposited over them to prevent chemical reaction between mercury and copper. A  DC current $I$ of a few Amp\`eres is applied through these electrodes , giving rise to a current density ${\bf j}$, and therefore a Lorentz force ${\bf F}_L={\bf j}\times{\bf B}$, as shown in Fig.~\ref{Fig1}. The surface is kept clean by maintaing the fluid in a nitrogen-filled atmosphere and is temperature-regulated by circulating water at 20.0 $\pm$ 0.1$^{\circ}$C. An electromagnetic vibration exciter, driven by a frequency synthesizer, provides a clean vertical sinusoidal acceleration (horizontal acceleration less than 1 $\%$ of the vertical one). The effective gravity in the reference frame of the container is then $g_{eff}(t)=g+acos(2\pi f_{ex} t)$, where $g$ is the acceleration of gravity, $a$ is proportional to the applied tension $V$ with a 1.0 Vs$^{2}$/m sensitivity and $f_{ex}$ is the excitation frequency. The sinudoidal modulation of $g$ is measured by a piezoelectric accelerometer and a charge amplifier. The surface wave amplitude is measured by two inductive sensors (eddy-current linear displacement gauge, Electro 4953 sensors with EMD1053 DC power supply). Both sensors, 3 mm in diameter, are screwed in the Plexiglas plate perpendicularly to the fluid surface at rest. They are put 0.7 mm above the surface on one diagonal of the cell, each one 15 mm away from its center (see Fig.~\ref{Fig1}). The linear sensing range of the sensors allows distance measurements from the sensor head to the fluid surface up to 1.27 mm with a 7.9 V/mm sensitivity. The linear response of these inductive sensors in the case of a wavy liquid metal surface has been checked in a previous study \cite{Precursor}.

In addition, local velocity fluctuations of the flow are measured by means of two Viv\`es probes\cite{Vives}, on opposite walls of the container. They are placed at 15 mm and 30 mm from the closest walls and separated from each other by 50 mm, as shown in Fig.~\ref{Fig1}. Each probe is made up by two copper wire electrodes plunging 2 mm into the fluid, separated by a distance $l$=3 mm and isolated completely from the liquid metal, except at the very end, where the electrical contact is made. A small cylindrical magnet ($\phi$=5 mm) is placed 5 mm above the electrodes, generating a magnetic field strength of $500$ G at the electrical contact points. The whole system is integrated into a cylindrical rod that is screwed to the Plexiglass plate. For velocity fluctuations of length scales larger than $l$, the voltage difference measured between the electrodes is proportional to $v_{1}B_{0}l$, where $v_1$ are the velocity fluctuations ortogonal to the vertical magnetic field $B_{0}$ generated by the probe's magnet\cite{Vives, Berhanu}. For scales much smaller than $l$, the small-scale velocity fluctuations are spatially integrated by the probe. In frequency domain, this means the transfer function of the probe is constant up to a cut-off frequency $f_c=\sigma(v_1)/2\pi l$, where $\sigma(v_1)$ is the {\it rms} value of the local velocity fluctuations. For frequencies larger than $f_c$, the transfer function of the probe decreases as $f^{-2}$\cite{Berhanu}. The small voltage voltage difference of the order of a few microvolts is amplified by a factor of order $10^{5}$ and acquired with the local height fluctuations and acceleration signals. The DC component of the signals is eliminated in the acquisition. The  sampling frequency is fixed at 500 Hz in order to resolve the temporal fluctuations of the measured quantities and the acquisition time is 800 s, much larger than the typical time scales of the acquired signals.

\section{Experimental results}

We start by describing the flows generated when the two forcing mechanism are applied separately. First, we describe the properties of the local wave amplitude and velocity profile of the parametrically amplified surface waves. Then, we consider the spatially periodic electromagnetic forcing alone.

 \begin{figure}[h]
 \centering
 \includegraphics[width=0.45\textwidth]{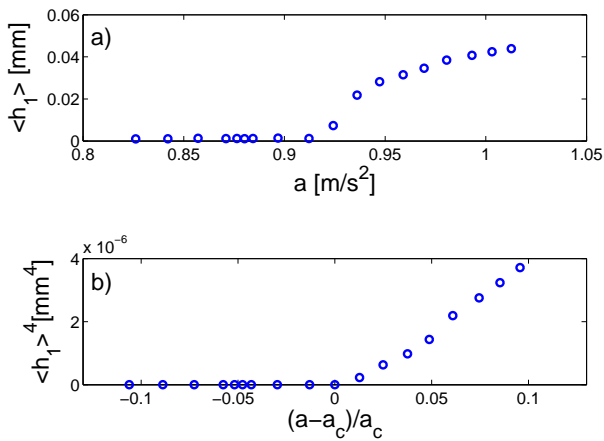}
 \caption{(a) Bifurcation diagram for the nonlinearly saturated wave amplitude $\left<h_1\right>$ as a function of $a$. (b) Bifurcation diagram for $\left<h_1\right>^{4}$ as a function of $\epsilon=(a-a_c)/a_c$.}
 \label{Fig2}
\end{figure}

\subsection{Parametric excitation}

The fluid container is vertically vibrated at frequency $f_{ex} = 23.8$ Hz and the amplitude $a$ is increased. At a given threshold $a_c$, the flat surface becomes unstable to small perturbations and stationary surface waves appear, generating a pattern that oscillates at half the forcing frequency $f_{ex}/2 = 11.9$ Hz. In this experimental configuration, the geometry of the standing pattern is made of squares with a wavelength $\lambda$ of order $7-8$ mm without any defect. At this wavelength, the surface waves are dominated by gravity.

The excitation frequency is chosen in order to have an eigenmode of standing surface waves over the container with a wavelength comparable to the one of the magnetic field ${\bf B}$. In the explored frequency range (20  $< f_{ex}<$ 30 Hz), the wavelength of parametrically amplified waves is roughly $8$ to $10$ times smaller that the size of the container. The frequency difference between two successive resonance tongues is about $1$ Hz. By tuning the excitation frequency within a $1$ Hz interval, it is easy to work in the vicinity of the minimum of a resonance tongue, i.e.,  without detunning between half the excitation frequency and the natural oscillation frequency of the surface waves. In this parameter range, the local amplitude of the surface waves $\left< h_1\right>$ does not present large scale modulations. The bifurcation diagram of the nonlinearly saturated wave amplitude is shown in Fig. \ref{Fig2}. Its dependence on the reduced control parameter $\epsilon=(a-a_c)/a_c$ is proportional to $\epsilon^{1/4}$, as it has been previously reported elsewhere\cite{Petrelis}. No distinguishable hysteresis loop is found in the computed bifurcation diagrams. To compute the bifurcation diagram, we have used the Fourier amplitude of the field at $f_{ex}/2$, by taking $$\left<h_1\right>=\lim_{T\rightarrow\infty}\left|\frac{1}{2T}\int_{-T}^{T}h_1(t)e^{\pi i f_{ex} t}dt\right|,$$ where $T$ is the acquisition time, much larger than the oscillation period $\pi/f_{ex}$ ($Tf_{ex}\sim 10^{4}$).

The surface wave dynamics is driven by the velocity field, which in turn is determined by the boundary condition given by the surface state. Hence, the velocity field also saturates nonlinearly. We have computed the bifurcation diagram of the local velocity field $v_1$, measured by the Viv\`es probes. The dependence on the reduced parameter $\epsilon$ of $v_1$ is proportional to $\epsilon^{1/2}$ (not shown here), contrary to the local amplitude dependence. This change in the saturation exponent can be explained by the nonlinear coupling of the subharmonic response of $v_1$ and the harmonic oscillation of the evanescent wave generated by the motion of the fluid's meniscus. In any case, both the local wave amplitude and velocity field present the same threshold value for $a_c$, showing the growth of one single mode over the container.

This weakly nonlinear regime, with a stationary nonlinearly saturated standing wave, will be studied when fluctuations in space and time are added to the wave system, through an underlying vortex flow. 

\subsection{Vortex flow}

We now study the fluctuations of the surface of the layer of mercury driven by a spatially periodic Lorentz force. When a current density ${\bf j}$ is applied through a liquid metal in the presence of a magnetic field ${\bf B}$, a Lorentz force density ${\bf F}_L={\bf j \times B}$ sets the fluid in motion. In the present configuration, the current density ${\bf j}$ is generated by a constant DC current $I$ applied through the mercury layer and its value is controlled externally by means of a power supply. The magnetic field ${\bf B}$, as explained above, is created by alternating polarity magnets arranged in an hexagonal lattice at the bottom of the container. Hence, ${\bf F}_L$ presents the same periodicity of ${\bf B}$ and generates a vortex flow which develops throughout the fluid pertubating the flat free surface and creating local height fluctuations. Small scale excitation using electromagnetic forcing has been used to study vortex dynamics\cite{kolmogorov,Tabeling} and quasi-two dimensional turbulence\cite{Sommeria}. The waves at the interface being of very small amplitude with respect to the depth of the mercury layer in our experimental setup up,  the current density ${\bf j}$ can be estimated as ${\bf j}=(I/S) {\bf e}$, where $S$=3.5 x 10$^{-2}$m$^{2}$ is the surface crossed by the current and ${\bf e}$ is a unitary vector pointing normally from one electrode (the cathode) to the other one (the anode). The velocity field ${\bf v}$ of the vortex flow can be estimated by balancing ${\bf F}_L$ and $\rho({\bf v}\cdot\bigtriangledown){\bf v}$ in the Navier-Stokes equation 
$$ \rho\left(\frac{\partial{\bf v}}{\partial t} + ({\bf v}\cdot\bigtriangledown){\bf v}\right)=-\bigtriangledown p +\rho\nu\bigtriangleup {\bf v} +{\bf F}_L,
$$ where $\rho$ is the fluid density and $\nu$ is its kinematic viscosity. The order of magnitude for such velocity fluctuations at the forcing scale (the wavelength $\lambda$ of the periodic magnetic field ${\bf B}$) for a current  $I$ of order $1$ A is 10$^{-2}$ m/s, thus giving a Reynolds number $Re$ of order 100. Even at low $Re$, the velocity field creates deformations on the free surface. Both surface and bulk fluctuations present large amplitude events and low-frequency fluctuations, as it is shown below. 

\begin{figure}[h]
 \centering
 \includegraphics[width=\columnwidth]{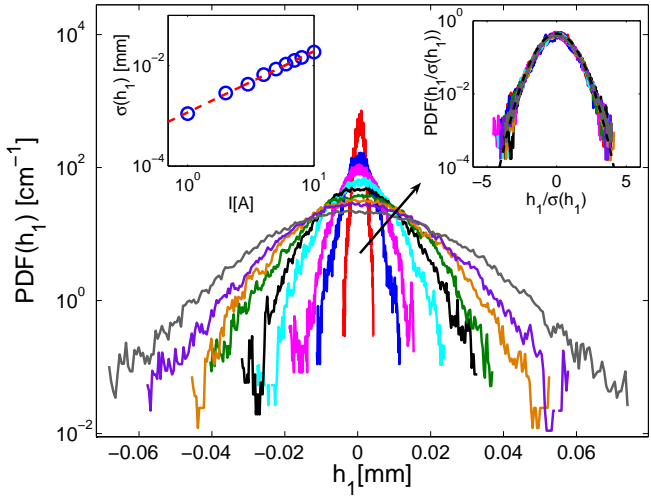}
 \caption{(color online) Probability density functions of the local amplitude fluctuations $h_1$ for different values of $I$ between 1 and 9 A. The arrow shows the sense of increasing current. Left inset: Loglog plot of $\sigma(h_1)$ as a function of $I$. The best fit slope is $1.2$. Right inset: Probability density functions of the rescaled local amplitude fluctuations $h_1/\sigma(h_1)$ for different values of $I$ between 1 and 9 A. The dashed line represents a gaussian fit.}
 \label{Fig3}
 \end{figure}

\subsubsection{Probability density functions}

\begin{figure}[h]
 \centering
 \includegraphics[width=\columnwidth]{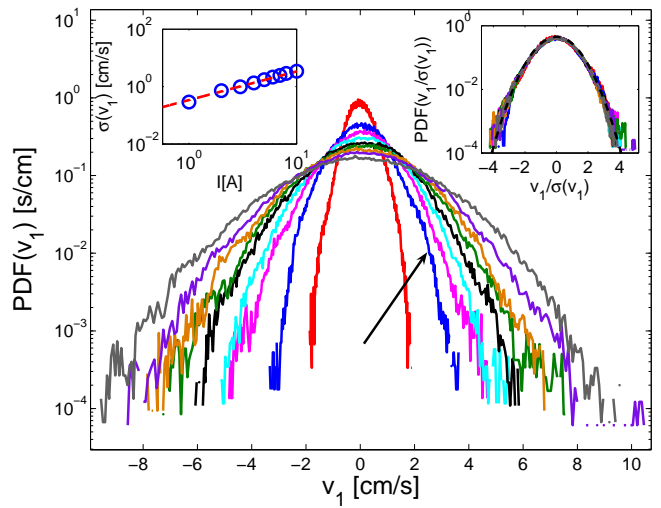}
 \caption{(color online) Probability density functions of the local velocity fluctuations $v_1$ for different values of the $I$ between 1 and 9 A. The arrow shows the sense of increasing current. Left inset: Loglog plot of the standard deviation $\sigma(v_1)$ as a function of $I$. Test fit slope is $1.0$. Right inset: Probability density functions of the rescaled local veocity fluctuations $v_1/\sigma(v_1)$ for different values of $I$ between 1 and 9 A. The dashed line represent a gaussian fit.}
 \label{Fig4}
 \end{figure}

To study the statistical properties of the local response of the fluid to the periodic Lorentz force, we compute the probability density function (PDF) of both the local surface amplitude $h_1$ and the velocity field fluctuations $v_1$.  We show their PDFs  in Fig. \ref{Fig3} and \ref{Fig4} for different values of the DC current $I$. Increasing the value of $I$, larger and larger fluctuating events of local height and velocity occur. The {\it rms} value of local surface fluctuations $\sigma(h_1)$ increases with increasing current, as does the {\it rms} value of the local velocity fluctuations $\sigma(v_1)$. Their dependence on $I$ is roughly linear (left inset in Fig. \ref{Fig3} and \ref{Fig4}). 

When plotted in the rescaled variables $h_1/\sigma(h_1)$ and $v_1/\sigma(v_1)$, all the PDFs collapse on one single curve (right inset in Fig. \ref{Fig3} and \ref{Fig4}). No clear asymmetry is found in the normalised PDFs of both variables. A slight departure from the statistics of a random gaussian variable is observed in both signals (the computed kurtosis is close to 3.2), but not large enough to discard gaussianity.
 
Contrary to the case of standing surface waves generated by parametric excitation, the local height and velocity fluctuations do not arise from the growth a single unstable mode. Hence, a certain degree of statistical independence between local height and velocity fluctuations appears when the electromagnetic forcing sets the fluid in motion. We can corroborate this fact by measuring their normalised covariance $$\rho_{x,y}=\frac{\overline{x(t)y(t)}}{\sigma(x)\sigma(y)},
$$where $x(t)$ and $y(t)$ stand for the local wave height or the local velocity signals and $\overline{()}$ stands for time average. 
Increasing $I$, increases the normalised covariance of the local wave height measured at two different points ($h_1$ and $h_2$) from  0.1 at 1 A till 0.25 at 8 A. In contrast, the normalised covariance for $v_1$ and $v_2$ fluctuates slighlty arround 0.1 for any value of $I$, the one of the pair of $v_1$ and $h_1$ behaves in a similar way. In that sense, the correlation lengths of the local height and velocity fluctuations decrease as the vortex flow intensity grows with $I$.

\subsubsection{Power spectral densities}

\begin{figure}[h]
 \centering
 \includegraphics[width=\columnwidth]{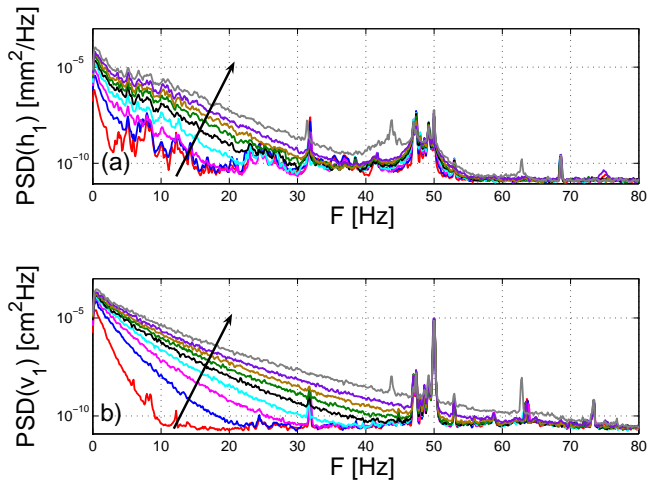}
 \caption{(color online) (a) Power spectral density the local wave amplitude fluctuations $h_1$ for different values of $I$ between 1 and 9 A as a function of the frequency $F$. (b) Power spectral density the local velocity field fluctuations $v_1$ for different values of $I$ between 1 and 9 A as a function of the frequency $F$. Arrows show the sense of increasing $I$.}
 \label{Fig5}
 \end{figure}

As $I$ is increased, low-frequency fluctuations dominate the response of the fluid motion to the Lorentz force ${\bf F}_L$. This is apparent in the power spectral densities (PSDs) of both $h_1$ and $v_1$ in shown Fig. \ref{Fig5}. The spectra of $h_1$ display an exponential behavior and no power-laws has been found even at large values of $I$ ($\sim$ 20 A, not shown here). This is not the case for the spectra of $v_1$ that neither have an exponential nor a defined power-law behavior. This corroborates the fact that even at the low values of $Re$ achieved in this experimental set-up, the flow remains strongly chaotic and fluctuating.

\begin{figure}[h]
 \centering
 \includegraphics[width=\columnwidth]{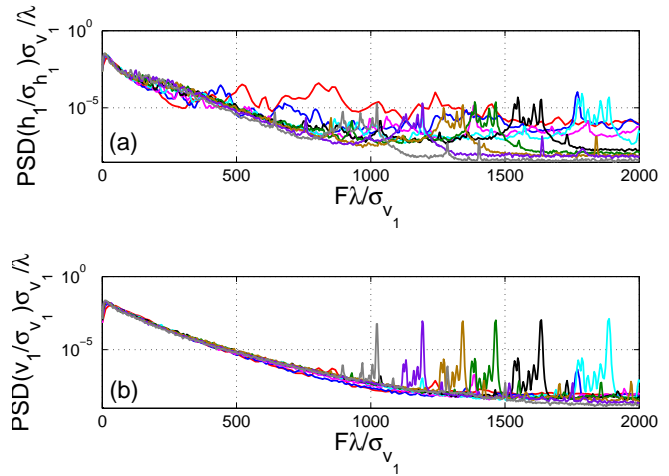}
 \caption{(color online) (a) Rescaled power spectral density of the normalised local wave amplitude fluctuations $h_1/\sigma(h_1)$ for different values of $I$ between 1 and 10 A as a function of the normalised frequency $F\lambda/\sigma(v_1)$. (b) Rescaled power spectral density of the normalised local velocity field fluctuations $v_1/\sigma(v_1)$ for different values of $I$ between 1 and 10 A as a function of the normalised frequency $F\lambda/\sigma(v_1)$.}
 \label{Fig6}
 \end{figure}

When $I$ is less than 1 A, there are clear peaks related to the lower normal modes of the surface waves in the container, excited by the fluctuations of the velocity field. At higher values of $I$, this coherent response is lost. Rescaling the frequency by the typical turn-over time of the vortex flow, $\lambda/\sigma(v_1)\sim$ 0.1 s,  and the PSDs of the normalised variables $h_1/\sigma(h_1)$ and $v_1/\sigma(v_1)$ by their inverse frequency, we can try to collapse all data into one single curve, as shown in Fig. \ref{Fig6}. For the local wave amplitude fluctuations, in the explored current range (1 A $< I <$ 10 A), there is a large dispersion for small values of $I$, due to the persistence of the cavity modes. As stated above, this coherent response is lost once the forcing is large enough ($I\sim5$A). On the other hand, all the velocity spectra collapse on one single curve. 

\section{Effect of spatio-temporal fluctuations induced by the vortex flow on parametric surface waves}

\begin{figure}[b]
\centering
\includegraphics[width=\columnwidth]{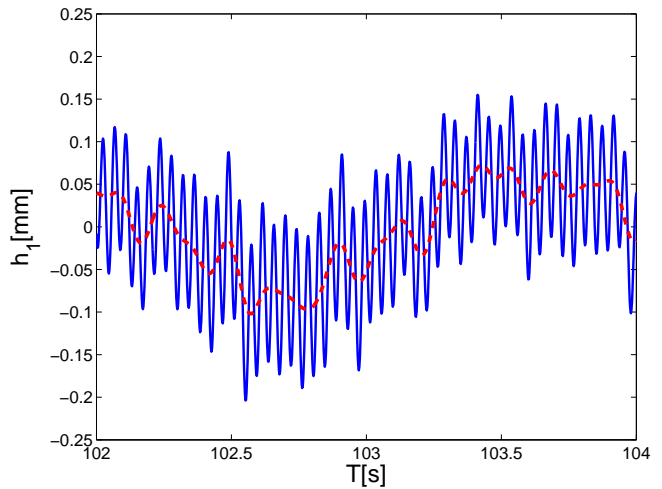}
\caption{(color online) Typical traces of the the local amplitude $h_1$ of the subharmonic response in presence of the vortex flow. The excitation frequency $f_{ex}$ is 23.8 Hz and the DC current $I$ is 1.0 A. Subharmonic oscillations (continous line) are observed over superimposed low-frequency fluctuations (dashed line), which are computed by low-pass filtering $h_1$ at 5 Hz.}
\label{FigTrazasruidoF}
\end{figure}

\begin{figure}[t]
 \centering
 \includegraphics[width=\columnwidth]{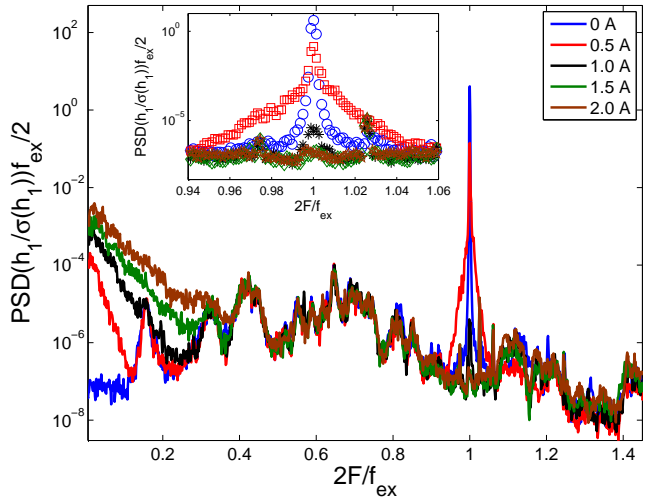}
 \caption{(color online) Power spectral density PSD the normalised local wave amplitude fluctuations $h_1/\sigma(h_1)$ for different values of the current intensity $I=$ 0 to 2 A as a function of the normalised frequency $2F/f_{ex}$ for a given value of $a>a_c$. Inset: Frequency window centered on the resonance peak of the parametric response for $I=$ 0 ($\circ$), 0.5 ($\square$), 1.0 ($\ast$), 1.5 ($\diamond$) and 2.0 ($\star$) A.}
 \label{Figh1}
 \end{figure}

Let us now study the effect of the velocity fluctuations driven by the periodic Lorentz force ${\bf F}_L$ on the growth, saturation and statitics of parametrically forced surface waves. The wavelength of the standing pattern is chosen to be of the same order of magnitude as the one of the periodic vortex flow, forced at wavelength $\lambda$. This is done to maximize the effect of the the vortex flow on the standing wave pattern. In the presence of the spatio-temporal fluctuations generated by the vortex flow, the local amplitude $h_1$ of the surface waves strongly fluctuates as shown in Fig. \ref{FigTrazasruidoF}. 
The PSD of $h_1$ is displayed in Fig. \ref{Figh1} for increasing values of the current $I$ that generates the vortex flow. We observe that the subharmonic response of the surface waves decreases when $I$ is increased. In contrast, the low frequency part of the spectrum increases. Correspondingly, the width of the subharmonic response is increased as shown in the inset of Fig. \ref{Figh1}. For larger currents, the amplitude of the subharmonic response keeps decreasing until it disapears under the noise level of the fluctutations generated by the Lorentz force. The parametric amplification of surface waves is thus inhibited by the spatio-temporal fluctuations generated by the vortex flow.
 
\begin{figure}[t]
 \centering
 \includegraphics[width=\columnwidth]{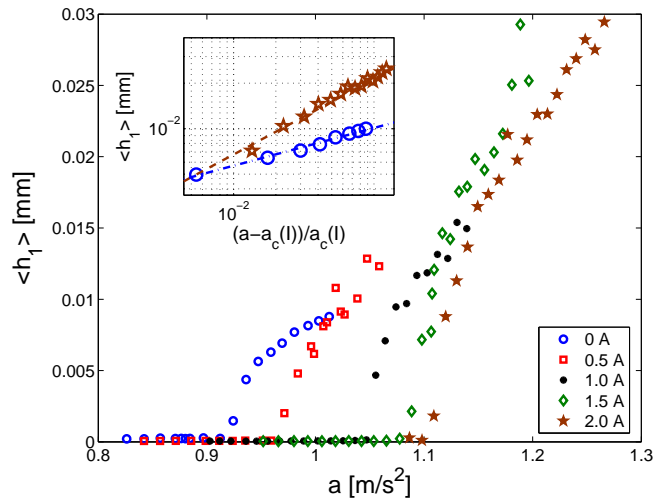}
 \caption{(color online) Bifurcation diagram of the local wave amplitude of the subharmonic response $\langle h_1\rangle$ for $I=$ 0 ($\circ$), 0.5 ($\square$), 1.0 ($\ast$), 1.5 ($\diamond$) and 2.0 ($\star$) A. Inset: Loglog plot of the local wave amplitude of the subharmonic response $\langle h_1\rangle$ as a function of the normalised acceleration $(a-a_c(I))/a_c(I)$ for $I=0$ A ($\circ$) and $I=2.0$ A ($\star$). Two power laws (dashed lines) are displayed with slopes 1/4 for $I=0$ A and 1/2 for $I=2.0$ A. }
 \label{Fig8}
 \end{figure}

In other words, the threshold of the parametric instability, i.e.,  the critical acceleration $a_c(I)$ for the onset of the subharmonic response, shifts to higher values with increasing values of $I$, as shown in the bifurcation diagram of the local wave amplitude (see Fig. \ref{Fig8}). We use the Fourier coefficients at $f_{ex}/2$ to compute the bifurcation diagrams as described above. No distinguishable hysteresis loop is found in the bifurcation diagram of the local wave amplitude. At a given value of the reduced control parameter $\epsilon(I)=(a-a_c(I))/a_c(I)$, the nonlinearly saturated local wave amplitude increases with $I$. The dependence on the dimensionless parameter $\epsilon(I)$ for $\left< h_1\right>$ changes as the vortex intensity increases: the 1/4 exponent for purely parametrically excited waves changes to 1/2 when both parametric and electromagnetic forcings are exciting the fluid motion (see the inset of Fig. \ref{Fig8}). This is a strong indication of the modification of the saturation mechanism of parametrically amplified surface waves induced by the nonlinear interaction between both the cellular flow of parametrically amplified waves and the underlying vortex flow. 

\begin{figure}[h]
 \centering
 \includegraphics[width=\columnwidth]{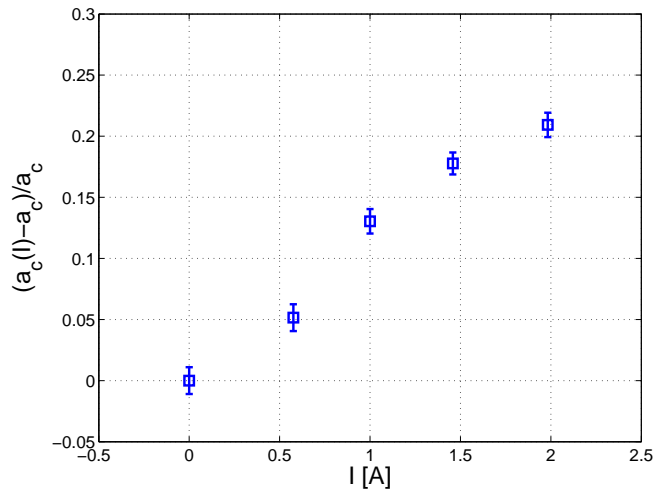}
 \caption{Normalised threshold growth $(a_c(I)-a_c)/a_c$ as a function of $I$ ($\square$).}
 \label{FigOnset}
 \end{figure}

From these bifurcation diagrams, we compute the threshold value $a_c(I)$ as a function of $I$ in the range 0 $<I<$ 2.0 A, as shown in Fig. \ref{FigOnset}. Increasing further the value of $I$ makes the fluctuations of the vortex flow comparable to the subharmonic response of parametrically amplified waves. Then, we cannot consider the underlying vortex flow as spatio-temporal perturbation of the wave pattern any more. This is noticeable from the inset of Fig. \ref{Figh1}, where the coherent subharmonic response is completely inhibited for large enough values of $I$. From Fig. \ref{FigOnset} we can see that the threshold value of the parametric instability can be shifted by 20$\%$ for $I=$ 2 A. 

Two simple mechanisms can be considered in order to explain the shift in threshold and the modification of the exponent from 1/4 to 1/2 of the scaling of the the parametric wave amplitude. First, random advection of the waves by the fluctuating vortices detunes the system away from parametric resonance. This leads to an increase of threshold. For instance, this type of mechanism explains the increase of threshold observed for parametric forcing in the presence of phase noise~\cite{Petrelis}. The modification of the exponent can be also ascribed to an effective detuning if it has the correct sign with respect to the nonlinear frequency correction.  However, it has not been observed in the presence of phase noise. Another explanation for both the frequency shift and the modification of the exponent can be based on an increased effective viscosity of the waves related to the underlying flow.  It is indeed known that the 1/4 exponent is observed at zero detuning only in the limit of small dissipation. It is likely that both mechanisms are involved, as for instance in the case where amplitude noise is added to parametric forcing. This indeed leads to both an effective detuning and dissipation~\cite{Stratonovich}.  
The determination of the relative contribution of these effects requires additional theoretical work.

We have also computed the normalised covariance for the subharmonic response of the local wave amplitude and local velocity field. Increasing $I$ for a given value of $a>a_c$ decreases the normalised covariance of the subharmonic response, decorrelating the fields at different places in the container. This effect is stronger between the local amplitude and the velocity fields. 
To emphasize this point, we have computed their spectral coherence $$C_{\widehat{x},\widehat{y}}(f)=\frac{|\overline{\widehat{x}(f) \widehat{y}(f)^{*}}|}{(\overline{|\widehat{x}(f)|^{2}}\overline{|\widehat{y}(f)|^{2}})^{1/2}},$$ where $\widehat{x}(f)$ and $\widehat{y}(f)$ stands for the Fourier transforms of $x$ and $y$ at frequency $f$ and $()^{*}$ stands for complex conjugate. This coefficient relates the possibility of two waves to produce interferences at a given frequency.  Increasing $I$, and thus the intensity of the vortex flow, for a given value of $a>a_c$, decreases the spectral coherence of the subharmonic response from 1.0 at 0 A till it reaches zero for $I$= 2.0 A, destroying the periodic nature of the oscillation of the parametrically excited waves. In that sense, the vortex flow prevents the standing wave pattern from maintaining its structure over the container.

\section{Conclusions}

     We have studied two experimental configurations related to the problem of wave-vortex interaction. We have first shown how an electromagnetically driven array of vortices in a layer of mercury generates surface waves and we have studied their statistical properties. We have observed that they qualitatively differ from the ones of surface waves generated by vibrating paddles~\cite{falcon}. Although broad band spectra of the local wave amplitude are easily obtained as soon as the underlying flow is spatio-temporally chaotic, they display an exponential cut-off instead of a power law decay observed for waves generated by vibrating paddles. The probability dentisy functions of local wave amplitude are also different: quasi-normal PDFs with standard deviations linearly increasing with the driving current are observed with electromagnetic forcing, whereas asymmetric PDFs have been reported in the case of vibrating paddles. We have also observed that the correlation between the waves and the underlying flow is small in the spatio-temporally chaotic regime. This has been also reported for waves generated by shear flows~\cite{vandewater}. In our experiment, the forcing of the waves by the array of vortices inhibits freely propagating waves. Thus, a regime of weak turbulence is not observed.
       
Second, we have shown that the underlying vortex flow acts as a source of spatio-temporal fluctuations that inhibits parametrically amplified surface waves. It modulates the amplitude of the subharmonic response and decorrelates the surface waves over the container, as shown by the measurement of the normalised covariance and spectral coherence of the subhamonic response: they both decrease when the underlying flow is strongly driven. The main effect of wave-vortex interactions is the growth of the threshold of the parametric instability. The vortex flow  also affects the saturation mechanism of the local wave amplitude and qualitatively modifies the scaling of the amplitude of the parametric waves as a function of the reduced control parameter $\epsilon(I)$. In the presence of the vortex flow $\langle h_1 \rangle\sim\epsilon(I)^{1/2}$, whereas $\langle h_1 \rangle\sim\epsilon(I=0)^{1/4}$ for parametrically amplified surface waves without forcing by the vortices. The theoretical description of qualitative and quantitative changes in the bifurcation diagram of the wave amplitude is needed. Work in that direction is in progress.
   
\section*{Acknowledgments}
The authors thank F. P\'etrelis for fruitful discussions over the subject. C. F. thanks R. Rojas and M. G. Clerc for corrections of the manuscript and acknowledges the financial aid of CONICYT and the constant support of Asefe. These experiments have been supported by CNES and ANR turbonde BLAN07-3-197846.

\end{document}